\documentclass{phb-proc4-auth}

\usepackage{graphicx}
\usepackage{amssymb}

\begin{document}
\begin{frontmatter}


\journal{SCES '04}


\title{Thermal conductivity of one-dimensional spin-$1/2$ systems}


\author[BS]{F. Heidrich-Meisner\corauthref{1}},
\author[BS]{A. Honecker},
\author[ST]{D.C. Cabra},
\author[BS]{W. Brenig}

 
\address[BS]{Institut f\"ur Theoretische Physik, Technische Universit\"at Braunschweig,  Germany}
\address[ST]{Laboratoire de Physique Th\'{e}orique, Universit\'{e} Louis Pasteur, Strasbourg, France}


%

\corauth[1]{Corresponding Author:  Institut f\"ur Theoretische Physik, TU Braunschweig,
Mendelssohnstrasse 3, 38106 Braunschweig, Germany. 
Phone: +49-(0)531-391-5184,
Fax: +49-(0)531-391-5833,
Email:f.heidrich-meisner@tu-bs.de
}


\begin{abstract}
We analyze the thermal conductivity of frustrated spin-$1/2$ chains within linear-response theory 
focusing on its frequency dependence at finite temperatures. Using exact diagonalization, 
the low-frequency limit
of  the thermal conductivity is discussed in the high-temperature regime.
\end{abstract}

%



\end{frontmatter}

%

The recent interest in    thermal transport properties of one-dimensional spin-$1/2$ systems
is motivated by the experimental observation
of significant contributions to the thermal conductivity of quasi-one dimensional materials originating from magnetic 
excitations  (see, e.g., Refs.~\cite{sologubenko00,hess01} and further examples mentioned in Ref.~\cite{hm03}).
A number of recent theoretical papers have either focused on the integrable $XXZ$
model \cite{hm03,zotos97,kluemper02,hm02,hm04}  or
on generic spin models like frustrated chains or spin
ladders \cite{hm03,hm02,alvarez02,orignac02,saito03,shimshoni03,zotos04}.
When analyzing transport properties, the real part of the conductivity is usually decomposed into a singular part
at zero frequency $\omega=0$ with weight $D_{\mathrm{th}}$ and a regular part
\begin{equation}
\mbox{Re}\,\kappa(\omega) = D_{\mathrm{th}}(T) \delta(\omega) +\kappa_{\mathrm{reg}}(\omega) \,.
\label{eq:1}
\end{equation}
Here, $\kappa$ denotes the thermal conductivity,  $D_{\mathrm{th}}$ is the thermal Drude weight,  
$T$ is the temperature, and $\delta(\omega)$ is the $\delta$-function.\\
\indent
The $XXZ$ model shows ballistic thermal transport properties due to the exact conservation
of the energy-current operator \cite{zotos97}. A complete understanding of the temperature and magnetic
field dependence of the relevant quantity, the thermal Drude weight $D_{\mathrm{th}}$, has recently
been established \cite{hm03,kluemper02,hm02,hm04}.
\\
\indent 
Regarding spin ladders and frustrated chains,
 both numerical \cite{hm03,hm02,zotos04} and field theoretical studies
\cite{hm03,shimshoni03} indicate  normal transport behavior,
characterized by a vanishing thermal Drude weight in the thermodynamic limit. 
Thus, the information relevant for the interpretation of experiments is encoded in the 
low-frequency behavior of the thermal conductivity. 
\\\indent 
This paper  focuses on the example of the frustrated
chain in the massive regime (see Ref.~\cite{zotos04} for a similar study of spin ladders). 
The Hamiltonian reads 
\begin{equation}
H= \sum_{l=1}^{N}h_l=  J\sum_{l=1}^{N} \lbrack  \vec{S}_l\cdot \vec{S}_{l+1} + \alpha\,  \vec{S}_l \cdot \vec{S}_{l+2}
\rbrack \,.
\label{eq:2}
\end{equation}
$h_l$ is the local energy density; $ \vec{S}_l$ is a spin-$1/2$ operator acting on site $l$.
$N$ denotes the number of sites and periodic boundary conditions are imposed. $\alpha J$ is the next-nearest
neighbor coupling; for $\alpha=0$, we obtain the spin-$1/2$ Heisenberg chain. In the following, we will
consider the case  of $\alpha=1$.
The  energy-current operator corresponding to this Hamiltonian is chosen such that 
the equation of continuity $-\partial_t h_l = \mbox{div}\, j_{\mathrm{th},l}$ is fulfilled \cite{hm02}
\begin{equation}
j_{\mathrm{th}}=i\sum_{l=1}^{N}j_{\mathrm{th},l}= i\sum_{l=1}^{N}\sum_{r,s=0}^{1} \lbrack h_{l-r-1},h_{l+s} \rbrack\,.
\label{eq:3}
\end{equation}
Following linear-response theory, the thermal conductivity can be written  as \cite{mahan}
\begin{equation}
\kappa_{\mathrm{reg}}(\omega) =  \frac{\pi\,g(\omega)}{N\,T}  \sum_{E_n\not=E_m}
\hspace{-0.15cm} p_n \,|\langle m |j_{\mathrm{th}} | n \rangle|^2 \delta(\omega-\Delta E)
\label{eq:4}
\end{equation}
with  $p_n=\mbox{exp}(- E_n/T)/Z$ being the Boltzmann weight, $Z$ the partition function, $\Delta E=E_m-E_n$, and
$g(\omega)=\lbrack 1-e^{-\omega/T}\rbrack /{\omega}$. $E_n$ are eigen-energies and $|n\rangle $ eigen-states
of the Hamiltonian Eq.~(\ref{eq:2}). Another useful quantity is the integral of $\mbox{Re}\,\kappa(\omega)$
over frequency $\omega$
\begin{equation}
I(\omega) = D_{\mathrm{th}}(T) + 2\, \int_{0^+}^{\omega} d \omega' \, \kappa_{\mathrm{reg}}(\omega')
; \enspace I_0= \lim_{\omega\to\infty} I(\omega)\,.
\label{eq:5}
\end{equation}
Applying standard numerical techniques allows us to study systems with $N\leq 20$ sites.  The dimension of the largest subspace 
 is $~9250$ after exploiting
translational invariance and conservation of total $S^z$. \\
  \begin{figure}[t]
     \centering
     \includegraphics[width=\columnwidth]{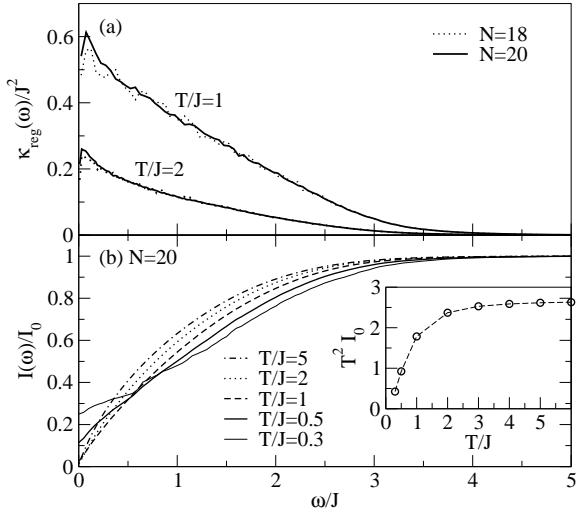}
     \caption{Thermal conductivity of  frustrated chains with $\alpha=1$ [see Eq.~(\ref{eq:1})].
     Panel (a): regular part of
     Re$\,\kappa(\omega)$ vs $\omega$ for $T/J=1,2$ for $N=18$ and $N=20$ sites.  Panel (b): 
     integrated weight $I(\omega)$ 
     [see Eq.~(\ref{eq:5})] vs $\omega$ for various
     temperatures. Inset of (b): $T^2\cdot I_0$ vs temperature $T$.}
     \label{fig:1} 
 \end{figure}  
 Let us now turn to the discussion of the results presented in Fig.~\ref{fig:1}. Panel (a) contains
 the regular part $\kappa_{\mathrm{reg}}(\omega)$ for $T/J=1,2$ and $N=18$ and $N=20$. 
 In this plot, $\kappa_{\mathrm{reg}}(\omega)$ has been binned with $\Delta E/J \sim 10^{-2}$.\\
 \indent
 First, note that $\kappa_{\mathrm{reg}}(\omega)$ is a structureless function of frequency $\omega$.
 Depending on temperature, the data are well converged down to frequencies $\omega/J \lesssim 0.25$ while
 deviations between $N=18$ and $N=20$ sites become visible for lower frequencies. Furthermore, we  observe a characteristic down-turn
 of $\kappa_{\mathrm{reg}}(\omega)$
 at low frequencies which is also present for spin ladders \cite{zotos04}. 
 This feature  is less pronounced on larger systems  because spectral weight is transferred from 
 the Drude weight to finite frequencies as the system size increases. Moreover,
 we have to keep in mind that the Drude weight is  excluded in the curves displayed in Fig.~\ref{fig:1}.
 Note that  the Drude weight amounts to less than 3\% of  the total weight $I_0$ from Eq.~(\ref{eq:5}) 
 for $N=20$ sites and $T/J=1$. Thus, the scenario of ballistic thermal transport 
 is unlikely for frustrated chains.  We refer the reader to 
 Refs.~\cite{hm03,hm02} for a detailed scaling analysis of the Drude weight.  
\indent
While the down-turn at low frequencies renders it difficult 
 to draw precise conclusions about the functional form of the frequency dependence,
 we can, however, fairly well estimate the absolute value of $\kappa=\lim_{\omega\to
 0}\kappa_{\mathrm{reg}}(\omega)$. For instance, the data shown in Fig.~\ref{fig:1}~(a) 
 indicates $\kappa/J^2\approx (0.27\pm 0.3) $ for $T/J=2$.\\
 \indent
 Next we comment on the integrated spectral weight $I(\omega)$, Eq.~(\ref{eq:5}), which  is 
 shown as $I(\omega)/I_0$ in Fig.~\ref{fig:1}~(b) for $N=20$ sites and various temperatures $T/J\geq 0.3$.
 Here, we see that first, $I(\omega)/I_0$ approaches $1$ the faster the larger $T$ is. 
 Second, the curve for $T/J=0.3$ is less smooth than the others indicating the limitations of an 
 exact diagonalization study
 with respect to the accessible temperature range. A smooth curve $\mbox{Re}\,\kappa(\omega)$ is 
 obtained for $T/J>0.5$ for $N=20$. Finally, we mention that $I_0 \sim T^{-2}$ for high temperatures, as
 expected for thermal transport [see inset of Fig.~\ref{fig:1}~(b)].\\  
 \indent
 We acknowledge financial support by the Deutsche Forschungsgemeinschaft
 and we would like to thank the Rechenzentrum of the TU Braunschweig, where parts of the numerical 
 computations have been performed.
%

%


\end{document}